\documentclass[aps,twocolumn]{revtex4}

\usepackage{amsmath}
\usepackage{amssymb}
\usepackage{comment}
\usepackage{graphicx}
\usepackage{xcolor}

\usepackage{hyperref}
\usepackage{ulem}

\begin{document}

\title{Analytical insights from a model of opinion formation \\based on Persuasive Argument Theory}

\author{
Lucía Pedraza$^{1,2,3}$,
Nicolas Saintier$^{4,5}$,
Juan Pablo Pinasco$^{4,6}$,\\
Pablo Balenzuela$^{1,2}$,
Celia Anteneodo$^{7,8}$,
 }

\affiliation{
$^1$Universidad de Buenos Aires, 
FCEN-DF. Buenos Aires, Argentina.
\\
$^2$CONICET - Universidad de Buenos Aires, 
INFINA. Buenos Aires, Argentina.
\\
$^3$Universidad de Buenos Aires, 
FCEN-DC. Buenos Aires, Argentina.
\\
$^4$Universidad de Buenos Aires, 
FCEN-DM. Buenos Aires, Argentina.
\\
$^5$CONICET - Universidad de Buenos Aires, 
IC. Buenos Aires, Argentina.
\\
$^6$CONICET - Universidad de Buenos Aires, 
IMAS. Buenos Aires, Argentina. 
\\
$^7$PUC-Rio, Departamento de Física \&  
$^8$INCT-SC. Rio de Janeiro, Brasil.
}


\begin{abstract}
In recent years, numerous mathematical models of opinion formation have been developed, incorporating diverse interaction mechanisms such as imitation and majority rule. However, limited attention has been given to models grounded in persuasive arguments theory (PAT), which describes how individuals may alter their opinions through the exchange of arguments during discussions. Moreover, analytical investigations of PAT-based models remain sparse. In this study, we propose an analytical model rooted in PAT, demonstrating that a group of agents can exhibit two distinct collective dynamics: quasi-consensus and bipolarization. Specifically, we explore various scenarios characterized by the number of arguments and the degree of homophily, revealing that bipolarization arises within this framework only in the presence of homophily.

\end{abstract}

\maketitle

\section{Introduction}

Societies frequently engage in debates that shape and evolve collective opinions on various issues. This dynamic process can lead to either consensus on specific topics or polarization, where opposing groups emerge with divergent views. A critical aspect of modeling these behaviors lies in defining the interaction mechanisms that underpin these phenomena. Some proposed mechanisms include imitation~\cite{akers1995social,bikhchandani1992theory}, persuasion~\cite{akers1995social, vinokur1978depolarization}, social pressure~\cite{festinger1950social,homans2017human} or kinetic exchanges~\cite{Crokidakis2012,Vieira2016}.

In this work, we explore a model in which the interaction mechanism follows the hypothesis of Persuasive Argument Theory (PAT) \cite{burnstein1973testing, burnstein1975person, laughlin1982social, mcguire1987group, mercier2011humans}. PAT is a proposed framework to explain how people change their opinions and behaviors. Unlike imitation- or comparison-based theories, PAT argues that when two individuals interact, they do not merely compare their opinions; instead they exchange a set of ideas, images, or thoughts related to the topic at hand, referred to as arguments. The theory assumes that there is a finite set of possible arguments, determined by cultural factors, and that when forming an opinion, individuals consider a subset of these arguments. The balance of arguments for or against a position influences both the direction and magnitude of the opinion. PAT provides an explanation for polarization phenomena, showing how a group initially in consensus can eventually reach a polarized state \cite{mas2013differentiation}.
It is our purpose to demonstrate that this effect can be achieved even with simplified interaction rules.

Opinion models can be analyzed using various approaches, such as agent-based models. However, deriving equations that approximate the dynamics provides a deeper understanding of the system, revealing the role of each parameter and helping to identify the transitions that lead to different states. Equations approximating the distance between opinions have been derived for several models in which the mechanism is based on imitation, whether discrete \cite{liggett1985interacting, sire1995coarsening, clifford1973model} or continuous \cite{DeGroot,Abelson}, with bounded confidence \cite{Toscani,Ye,Perez,Anteneodo}, in both one-dimensional \cite{ANT,PPS, PPSS,PSB,Vazquez,Tyson,paperopinionFitness2, Vazquez2, Vazquez3} and multidimensional opinion spaces \cite{Motsch, Noorazar,boudin,pedraza2021analytical}. However, no analytical studies have been performed on a model based on the PAT hypothesis.

In this study, we present a model based on PAT. Despite its simplicity, the model demonstrates complex dynamics, where the system may reach either consensus or polarization, depending on the parameters. Section 2 introduces the development of the model. By simplifying the framework such that each agent’s state is one-dimensional and discrete, we derive the master equations governing the dynamics, which depend solely on two parameters: the strength of homophily $\alpha$ and the number of arguments each agent possesses $M$. Section 3 details the derivation of the equations and validates them by comparing the results with numerical simulations. In Section 4, we explicitly solve the equations for some particular cases: $\alpha = 0$ (where all agents interact with equal probability) and for small values of $M$. Additionally, we obtain the continuous approximation of the equations for large $M$ and derive the equation in the limit as $M \to \infty$, concluding with the derivation of the final-state equation.

\section{Model}

We consider a group of $n$ agents. Each possesses a set of $M$ arguments. 
Arguments can only be positive ($+1$, in favor of the discussed issue) or negative ($-1$, against the discussed issue). Beyond their signs, the arguments are indistinguishable and have no specific order. The state of an agent is thus completely defined by the proportion of positive arguments $a$. Notice that the proportion of negative arguments is then $1-a$.

Each agent also holds an opinion, which is defined by the type of arguments they possess and is equal to the difference between the proportions of positive and negative arguments. Therefore, the opinion $o$ of an agent with a proportion of positive arguments equal to $a$ is $o=2a-1$. Since $a\in [0,1]$, the opinion ranges between $-1$ and $1$.

Agents can modify the composition of their arguments sets due to interactions with each other, and consequently modify their opinion. The interaction dynamics occurs in a sequence of events in which each agent can interact with and be influenced by another agent. The rate of interaction between two agents $i$ and $j$ is determined by their homophily defined as the similarity of their opinions $o_i$ and $o_j$ measured by 
\begin{equation}\label{hom2}
h(o_i,o_j) = (1-\frac12 \Delta o)^\alpha, \qquad \Delta o  = |o_i-o_j|,
\end{equation}

for a certain power constant $\alpha\ge 0$.  
The homophily can also be expressed in terms of their proportion of positive arguments as 
\begin{equation}\label{eq:hom}
h(a_i,a_j) = (1-\Delta a)^\alpha, \qquad \Delta a  = |a_i-a_j|. 
\end{equation}
Since the difference of opinion $\Delta o\in [0,2]$, the homophily ranges from 0 to 1. 
It is maximal, equal to 1, when both agents share the same opinion, i.e., have the same number of positive arguments,
and it is minimum, equal to 0, when they have completely opposite opinion, which occurs when one agent has only positive arguments, and the other only negative arguments. 
Notice that, when $\alpha=0$, the homophily becomes constant, equal to 1, so that the interaction rate between agents becomes independent of their opinion. When $\alpha>0$, the homophily depends nonlinearly on the difference of opinion, making the interaction between agents holding different opinion more and more unlikely as $\alpha$ increases.

\begin{figure}[h]
\includegraphics[width=8.5cm]{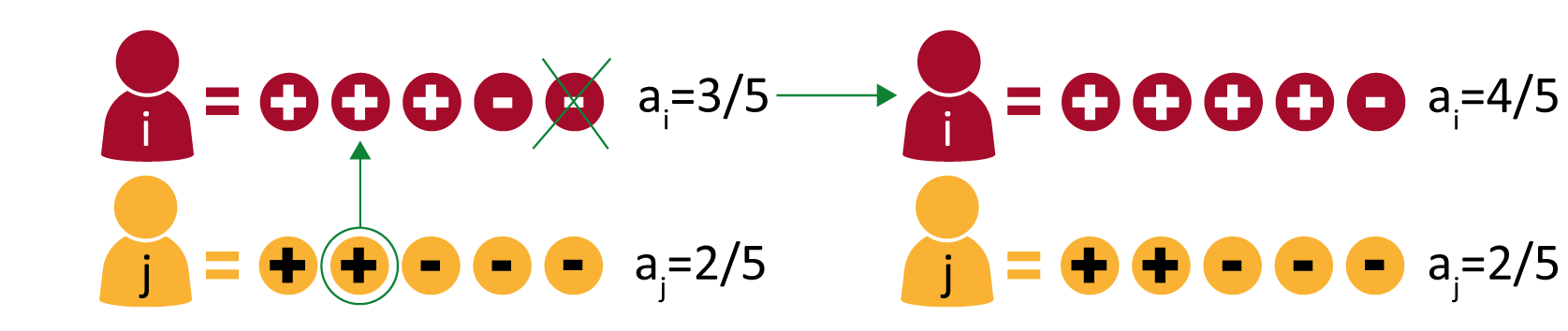}
\caption{Example of interaction for $M=5$. In this case, agent $i$ starts with 3 positive arguments and two negative ones, so $a_i=3/5$, and agent $j$ starts with 2 positive arguments and three negative ones, so $a_j=2/5$. With probability $h(a_i,a_j) = (\frac{4}{5})^\alpha$, the agents interact. In that case, agent $i$ randomly removes one of their own arguments (negative with probability $2/5$) and randomly takes one of agent $j$'s arguments, positive with probability $3/5$. Thus, in this case, agent $i$ updates their state to $4/5$.
}
\label{Fig:Esquema}
\end{figure}

At each time step, every agent $i$ is assigned another random agent $j$. With a probability $h(o_i, o_j)$, agent $i$ will be influenced by agent $j$. If this happens, the interaction will proceed as follows: 
\begin{itemize}
    \item First, agent $i$ will randomly discard one of its $M$ arguments, 
    \item then, $i$ will randomly take one of agent $j$'s $M$ arguments.
\end{itemize}
This process is sketched in Fig. \ref{Fig:Esquema}. 
The dynamics is repeated in the same way until the proportion of agents with each possible number of arguments remains stable.

\section{Theoretical analysis}

\subsection{Rate equations} 

When agent $i$ having a proportion $a_i$ of positive argument interacts with agent $j$ holding a proportion $a_j$ 
of positive argument, three situations can occur: 
\begin{itemize} 
\item $i$ loses a negative argument and gains a positive one from $j$. This occurs with probability $(1-a_i)a_j$, 
and in that case $a_i$ increases by $1/M$,
\item symmetrically, $i$ may gains a negative argument from $j$ and loses a positive one. 
This occurs with probability $(1-a_j)a_i$, 
and in that case $a_i$ decreases by $1/M$,
\item $i$ and $j$ both put forth arguments of the same sign resulting in $\Delta a_i=0$. 
\end{itemize} 
In summary 
\begin{equation}\label{eq:update_rule}
a_i \rightarrow 
\begin{cases}
a_i + \frac{1}{M} \qquad &\text{with probability } (1-a_i)a_j, \\
a_i - \frac{1}{M} \qquad &\text{with probability } a_i(1-a_j), \\
a_i \qquad  &\text{otherwise.} 
\end{cases}
\end{equation}

Let us denote $p_i(t)$ the expected proportion of agents with $i$ positive arguments, $i=0,..,M$, at time $t$.
Then $p_1,..,p_M$ satisfy the system of rate equations 
\begin{equation}\label{eq:RateEq}
\begin{split}
\frac{d p_i}{dt}  
 & =  p_{i+1}\sum_j p_j h(i+1,j)\frac{i+1}{M}\frac{M-j}{M} \\
 & + p_{i-1}\sum_j p_j h(i-1,j)\frac{M-(i-1)}{M}\frac{j}{M} \\ 
 & - p_i\sum_j p_j h(i,j) 
\Big( \frac{i}{M}\frac{M-j}{M} + \frac{M-i}{M}\frac{j}{M} \Big),
\end{split}
\end{equation}
where we let $p_{-1}=p_{M+1}=0$ and denote $h(i,j) :=h(i/M,j/M)$ the rate of interaction between two agents holding $i$ and $j$ positive arguments. 

The first two terms in the r.h.s. of Eq.~\eqref{eq:RateEq} correspond to the two possible ways of increasing the number of agents with $i$ positive arguments, namely through interaction between some agent holding $j$ positive arguments, $j=0,...,M$, with 
(i) either an agent with $i+1$ positive arguments who loses one positive argument and gains a negative one, 
(ii) or an agent with $i-1$ positive arguments who gains one positive argument and loses a negative one. 

The last term in the r.h.s. of Eq.~\eqref{eq:RateEq} corresponds to interactions between an agent holding $i$ positive arguments, either losing or gaining a positive argument during the interaction with some agent with $j=0,..,M$ positive arguments. 

Notice that the total mass $ \sum_j  p_j$ (which is equal to 1) and the mean proportion $\mu = \sum_j \frac{j}{M}p_j$ 
of positive arguments are conserved quantities. 
In fact, by summing the rate equations, one obtains $\sum_ip_i'=0$. 
Furthermore, summing each rate equations times $i/M$ yields, after changing variables $i\pm 1\to i$ in the first two sums, the following result:  
$$ M^3 \frac{d}{dt} \mu = \sum_{i,j} h(i,j)p_ip_j [j(M-i)-i(M-j)]=0, $$
since $h$ is symmetric, i.e., $h(i,j)=h(j,i)$.
Eventually, the symmetry $h(M-i,M-j)=h(i,j)$ implies that if the distribution $(p_0,...,p_M)$ is initially symmetric, 
i.e. $p_i=p_{M-i}$ for all $i$, it will remain symmetric for all subsequent times. 
This follows by noticing that the distribution $(\tilde p_0,...,\tilde p_M)$, where $\tilde p_i:=p_{M-i}$ satisfies the same system of equations as $(p_0,...,p_M)$.

Let us compare the results of the agent-based simulation with the solution of Eq.~\eqref{eq:RateEq}. 
In the simulations, we considered 1000 agents whose opinions are represented by a binary argument vector, where 1 denotes a positive argument and -1 denotes a negative one. Initially, each argument is randomly assigned a positive or negative value with a probability of $\frac12$. At each interaction, each agent is paired with another and interacs with a probability determined by their homophily, as defined in Eq.~(\ref{eq:hom}). If they interact, the first agent adopts a randomly selected argument from their partner and discards one of their own arguments, chosen at random.

\begin{figure}[t!]
\includegraphics[width=0.5\textwidth]{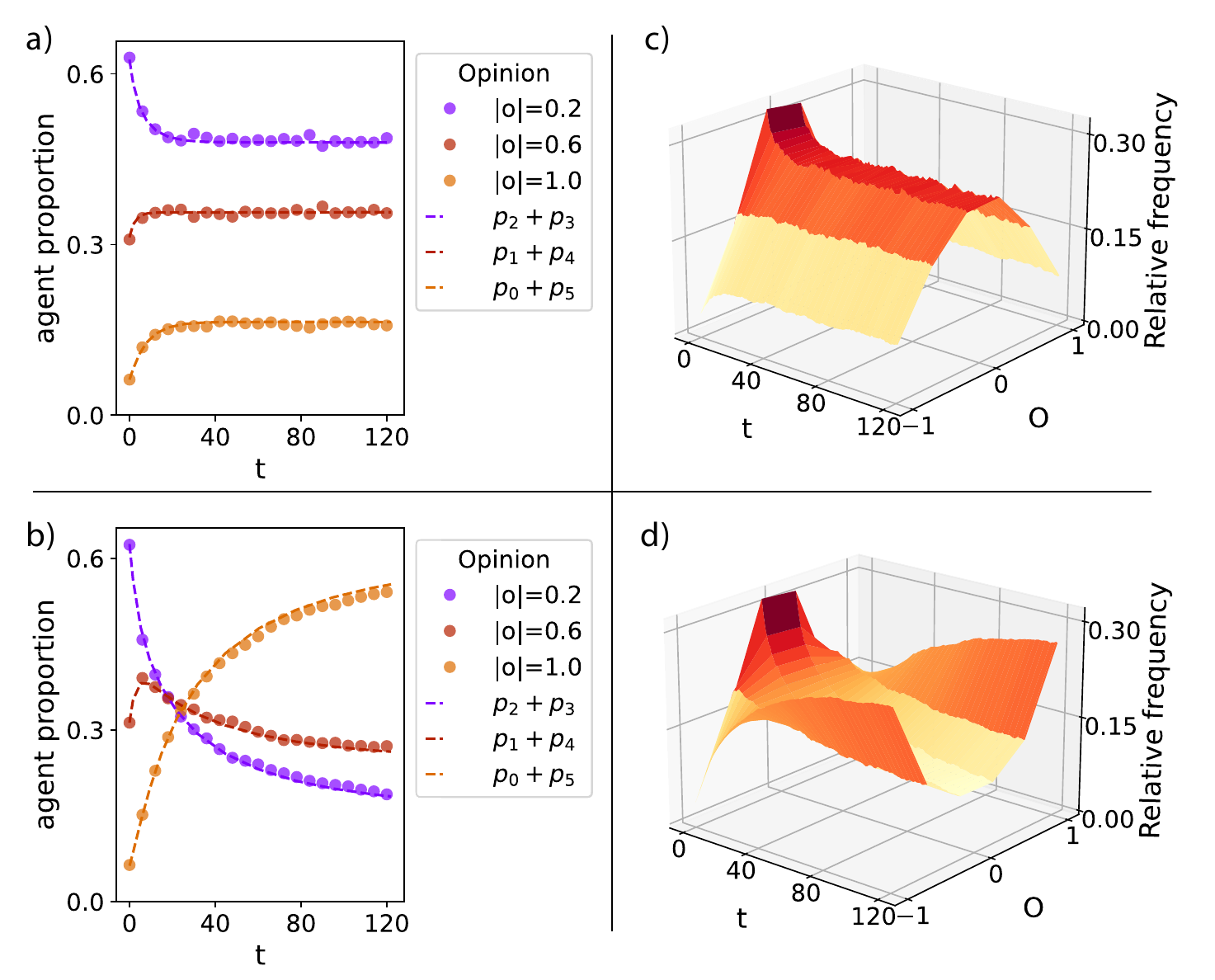}
\caption{The temporal evolution for $M=5$ is shown. Panels a) and c) display the proportion of agents by state, for $\alpha=1$, combining symmetric states, as obtained from simulations with 1000 agents (circles in a), and the numerical solution of Eq.~\ref{eq:RateEq} ( lines). 
Panels c) and d) show the temporal evolution of the agents' proportion in each state, for $\alpha=3$. 
In panels a) and c), for $\alpha=1$, a consensus process is observed, while panels b) and d), for $\alpha=3$, the resulting in a final state of bipolarization.
} 
\label{Fig:Evol_temporal}
\end{figure}

In Fig.~\ref{Fig:Evol_temporal}, we observe the evolution of the dynamics for the case of $M=5$ arguments and homophily exponent $\alpha=1$ (panels a and c) and $\alpha=3$ (panels b and d). Panels a and b show the proportion of agents over time, with the results from simulations represented by symbols, and the analytical results, obtained solving the rate equations \eqref{eq:RateEq}, represented by lines. Considering a symmetric distribution of agents, those with opinions of equal magnitude are grouped together. Panels c and d depict the temporal evolution of the proportion of agents with each opinion.

In these examples, two distinct final states can be observed. 
For the case of homophily exponent $\alpha=1$, the evolution stabilizes in a state where most agents hold neutral opinions. However, for the case where the homophily is stronger, with exponent $\alpha=3$, we find a bipolarized state where the majority of agents adopts extreme opinions.

Now, we aim to characterize the set of parameters for which the system reaches each of the possible states.
Since we found very good agreement between the agent-based simulations and the numerical resolutions of the rate equations, we can thus rely on the rate equations to study the impact of $M$ and $\alpha$ on the bipolarization of the final state.
In order to quantitatively assess the impact of these parameters on the final state of the system, we consider an index quantifying the bipolarization of a distribution $(p_0,..,p_M)$.
This is the so-called Bimodality Coefficient (BC), which for the distribution $P(X)$ (in our case $P(X=i/M)=p_i$) is defined as follows 
\begin{equation}
    BC = \frac{\text{skewness}^2+1}{\text{kurtosis}},
\end{equation} 
where the skewness and kurtosis are the standardized 3rd and 4th moments of $X$, i.e., 
$E[((X-\mu)/\sigma)^n]$, $n=3, 4$, where $\mu$ and $\sigma^2$ are the expected value and variance of $X$. 
The skewness is a measure of the asymmetry of $X$ (in particular it is 0 if $X$ is symmetric with respect to $\mu$),
and the skewness is a measure of tailedness. 
The definition of the BC index is motivated by Pearson's inequality 
$ {\text{kurtosis} \ge \text{skewness}^2}+1$ which gives $BC\le 1$ (see \cite{P}). 
The rule of thumb is that the closer BC is to 1 (resp. 0), 
the more bimodal (resp. unimodal) is $X$ - the threshold being $0.55$ which is the BC of a uniform 
distribution. 
However there are examples where this rule of thumb goes off (see \cite{Pfister, Tarba}) though these examples 
are mostly non-symmetric distributions. 
For example, the BC index of the final state of the case shown in Fig.~\ref{Fig:Evol_temporal}(a) is $0.46$ (unimodal state), while for the case shown in Fig.~\ref{Fig:Evol_temporal}(b), the BC index is $0.72$ (bimodal state).

\medskip

\subsection{Numerical solutions of rate equations.}

\begin{figure}[b!]
\includegraphics[width=0.48\textwidth]{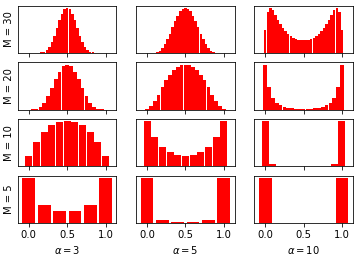}
(a) Symmetric ($\mu=0.5$)\\
\includegraphics[width=0.48\textwidth]{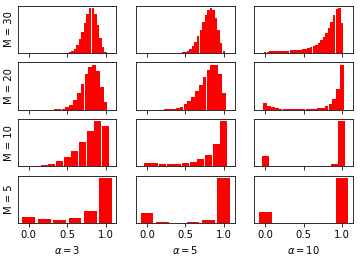}
(b) Asymmetric ($\mu=0.8$)
\caption{Stationary state of the rate equations (red) for $M=5,10,20,30$ (bottom to top)
and $\alpha=3,5,10$ (left to right). a) Symmetric case with mean value $\mu=0.5$. b) 
Asymmetric case with mean value $\mu=0.8$.
}
\label{fig:Rate_eq}
\end{figure}

Figure \ref{fig:Rate_eq} shows the final state resulting from the numerical solution of the rate equations for selected values of $M$ and $\alpha$. We consider two scenarios: one with a symmetric initial condition, where the mean value is $m = 0.5$ (a), and another with an asymmetric initial condition, where the mean value is $m = 0.8$ (b). In both cases, we observe the transition between unimodal and bimodal distributions.

\begin{figure}[ht!]
\includegraphics[width=0.45\textwidth]{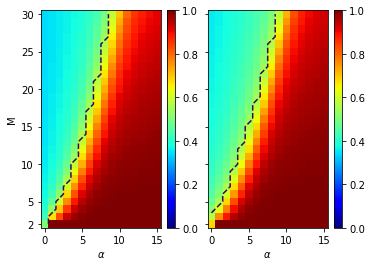}
\caption{Heatmap in the plane of parameters $\alpha$--$M$ of the degree of bimodality BC of the stationary state of the dynamic computed with the rate equations.
Symmetric case with mean value $\mu=0.5$ (left) and an asymmetric case with mean value $\mu=0.8$ (right).
 The threshold between unimodal and bimodal state is indicated by the dashed line. 
}
\label{Fig:barrido}
\end{figure}

To gain a clearer understanding of the transition between uni- and bimodal shapes as a function of $M$ and $\alpha$, we present in Fig.~\ref{Fig:barrido} a heatmap of the BC indices corresponding to the final stationary state of the rate equations, across a sweep of both the parameters, for the symmetric (a) and asymmetric cases (b). The dashed line divides the $M-\alpha$ plane into regions representing uni- and bimodal behavior, with the threshold 
$BC> 0.55$ demarcating the boundary, as discussed above.

We note that for $\alpha=0$, which corresponds to a constant homophily equal to 1,
the final state remains unimodal for any value of $M$. 
In the following section, we provide an analytical demonstration that the stationary state is described by a binomial distribution, which, for large values of $M$, can be approximated by a Gaussian distribution along with two Dirac masses.  

When $\alpha>0$, the final state is either bipolarized or unimodal, depending on whether $M$ is smaller or larger than acritical value. 
The transition between unimodal and bimodal states can be determined analytically for small values of $M$ by explicitly solving the rate equations, which we will elaborate in Sec.~\ref{sec:low}. 
For large values of $M$, we will instead use, in Sec.~\ref{sec:large}, a continuous approximation of the rate equation to analyze the transition.

Before proceeding, note that $(1,0,...,0)$ and $(0,..,0,1)$ are trivial stationary states of the rate equations, 
corresponding to $\mu=0$ and $\mu=1$, respectively. Therefore, we will assume that $\mu$ lies in $(0,1)$ in the following discussion.

\section{Analytical stationary solutions}
\label{sec:analytical-solutions}

In this section, we study analytically the transition observed in Fig.~\ref{Fig:barrido}. 
Although obtaining a general solution is not possible, we can explore it in specific scenarios: with and without homophily. In the former case, we analyze the particular case of small values of $M$ on the one hand, and by applying a continuous approximation for large values of $M$ for the other.

\subsection{Without homophily}

When $\alpha=0$, i.e., the homophily is constantly equal to 1, the rate equations in Eq.~\eqref{eq:RateEq} become
\begin{eqnarray*} 
p_i' 
& = & \frac{i+1}{M}p_{i+1}\sum_j (1-\frac{j}{M})p_j 
+ (1-\frac{i-1}{M})p_{i-1}\sum_j \frac{j}{M}p_j \\
&& - \frac{i}{M}p_i\sum_j (1-\frac{j}{M})p_j 
- (1-\frac{i}{M})p_i\sum_j \frac{j}{M}p_j,
\end{eqnarray*} 
with $p_{-1}=p_{M+1}=0$. 
Recalling that the total mass is $\sum_j p_j=1$ and that 
the mean proportion $\mu = \sum_j \frac{j}{M}p_j$ 
of positive arguments is also conserved, we obtain 
\begin{eqnarray*} 
p_i' & = & \frac{i+1}{M}(1-\mu) p_{i+1}
+ (1-\frac{i-1}{M})\mu p_{i-1} \\
&& - (\frac{i}{M} + (1-\frac{2i}{M})\mu)p_i. 
\end{eqnarray*} 
Thus, the $p_i$ solves a linear system of ordinary differential equations (a non-symmetric random walk). 
For the stationary state, we can find an explicit formula as follows. Equating the l.h.s to 0 and solving successively for $i=0, 1, ...$, yields $p_1, p_2, ...$ in terms of $p_0$ resulting in 
\begin{equation} 
p_i = \binom{M}{i} \Big(\frac{\mu}{1-\mu}\Big)^i p_0, 
\qquad \mbox{for, $i=0, .., M$}. 
\end{equation}
Additionally, the normalization condition $\sum_i p_i=1$, gives $p_0=(1-\mu)^M$ so that 
\begin{equation}  p_i = \binom{M}{i} \mu^i  (1-\mu)^{M-i}
\qquad \mbox{for $i=0, .., M$},
\end{equation} 
which is the binomial distribution ${\rm Bin}(M, \mu)$.

We can find a simple Gaussian approximation for large $M$. 
We know that $p$ is the law of $\tilde X_M:=\frac1M X_M$ with $X_M\sim Bin(M, \mu)$. 
It follows that the distribution $f_{\tilde X_M}$ of $\tilde X_M$ is 
$f_{\tilde X_M}(x) = M f_{X_M}(Mx)$. 
In the limit $M\to +\infty$, by the Central Limit Theorem, $X_M$ is approximately Normal with mean $M\mu$ and variance 
$M\mu(1-\mu)$. 
Thus 
\begin{equation}\label{DefGaussianApprox}
 f_{\tilde X_M}(x) \approx 
g_M(x):=\sqrt{\frac{M}{2\pi\mu q}} 
\exp\,\Big( - \frac{M (x-\mu)^2}{2\mu q}\Big) \
\end{equation}
where $q:=1-\mu$.
Notice, however, that this distribution must be restricted to $[0,1]$. 
Dirac masses at $0$ and $1$ thus appear with weights $a_M, b_M$ approximately equal to 
$$ a_M = \int_{-\infty}^0 f_{\tilde X_M}(x)\,dx,\qquad b_M = \int_1^{+\infty} f_{\tilde X_M}(x)\,dx.$$ 
Using the well-known asymptotic 
$$ \int_x^{+\infty} e^{-t^2}\,dt = \int_{-\infty}^{-x} e^{-t^2}\,dt  
\approx \frac{1}{2x} e^{-x^2}, \qquad x\gg 1,$$
we have, after a change of variable and letting $x_M=\sqrt{\frac{M\mu}{2q}}$,
that 
\begin{equation}\label{DefDiracApprox1}
 a_M \approx \frac{1}{\sqrt\pi} \int_{-\infty}^{-x_M} e^{-t^2}\,dt 
 \approx \sqrt{\frac{q}{2\pi \mu M}} 
 e^{-\frac{\mu M}{2q}}.
\end{equation}
In the same way
\begin{equation}\label{DefDiracApprox2}
b_M \approx \sqrt{\frac{\mu}{2\pi q M}} 
 e^{-\frac{q M}{2\mu}}.
\end{equation}
Thus, in the limit $M\gg 1$ the stationary state of the rate equation is well approximated by the Gaussian $g_M$ defined in Eq.~\eqref{DefGaussianApprox} but restricted to $[0,1]$
plus two Dirac masses at 0 and 1 with weights $a_M, b_M$ given approximately by Eqs~\eqref{DefDiracApprox1}-\eqref{DefDiracApprox2}.

\subsection{With homophily: Low values of $M$}
\label{sec:low}

We analyze the dynamics generated by the system of rate equations and its convergence to equilibrium for non-constant homophily ($\alpha>0$). 
The conservation of the mass and of the mean proportion $\mu$ of positive arguments allow to reduce the original system, given by Eq.~(\ref{eq:RateEq}), to a set of $M-1$ equations. In case of symmetric solutions, the number of equations can be further reduced, to 
$(M-1)/2$ (for odd $M$) or to $M/2$ for even $M$. 
In this section, we will show that, for $M=2$, the steady state becomes bimodal when $\alpha>0$, contrasting with the unimodal character of the homophily-less scenario ($\alpha=0$). 
For $M=3$ and $M=4$, there is typically a critical value of $\alpha$ that separates the unimodal regime (at low $\alpha$) from the bimodal one (at high $\alpha$). 
Then, a small number of arguments already manifests the behavior observed for very large $M$.

\subsubsection{ $M=2$}

For two arguments ($M=2$) and homophily ($\alpha>0$), we have $h(2,0)=h(0,2)=0$, $h(i,i)=1$ (for $i=1,2,3$), and $h(i,j)=2^{-\alpha}$ otherwise. From Eq.~(\ref{eq:RateEq}), and considering the normalization condition as well as the conservation of the mean $\mu \equiv \langle a \rangle=(p_1+2p_2)/2$, the evolution of the system can be described by a single equation, namely,
\begin{equation} 
    p_1'= -\frac12 p_1^2,
\end{equation}
which holds for any $\alpha>0$.
As time progresses, $p_1$ approaches zero, hence the system evolves towards the state $(1-\mu,0,\mu)$. 
Thus, the final state is bipolarized (being symmetric if $\mu=1/2$ and asymmetric for other values of $\mu$), except in the trivial cases when $\mu=0$ or $\mu=1$, which have already been excluded. 

Note that this picture contrasts with the case in the absence of homophily ($\alpha=0$), where the steady state is described by a binomial distribution, resulting in a unimodal profile. 
In this case, the dynamics is governed by the equation  
\begin{equation} 
    p_1'= 2\mu(1-\mu) - p_1,
\end{equation}
leading to the steady state
 $( (1-\mu)^2,2\mu(1-\mu),\mu^2)$.
This indicates that $p_1$ is maximal for $1/3<\mu<2/3$ (in particular in the symmetric case $\mu=1/2$) or one of the extremes is maximal for the mean outside this interval, except at the borderline values of $\mu$. For instance, when $\mu=2/3$, the final state is (1/9,4/9,4/9)].

\subsubsection{ $M=3$}

For three arguments, let us first analyze the symmetric case, where $p_3=p_0$ and $p_2=p_1$, yielding $\mu=1/2$. Additionally using the normalization condition, the dynamics can be described by a single equation, which depends on the degree of homophily, and for $\alpha>0$ is given by

\begin{equation} 
    p_1'= \frac{p_1}{3^{\alpha+1}}
    \bigl( 1- \frac{p_1}{c} \bigr),
\end{equation}
where $c=3/(2^\alpha + 6 + 2\;3^\alpha)$. This equation has two fixed points, namely $p_1=0$, which is unstable, and $p_1=c$, which is stable. 
Therefore, the system evolves to $(1/2-c,c,c,1/2-c)$, except when starting from the trivial (polarized) state $(1/2,0,0,1/2)$. 
Note that, the system is more bimodal than unimodal for $1/2-c>c$, which implies $\alpha > \alpha_{c} \simeq 0.7102$. 

For the general case, with arbitrary mean $\mu$, the dynamics is governed by the following set of two coupled equations,
\begin{eqnarray} \notag
    p'_0&=&
\frac{2\,3^\alpha - 2^{\alpha+1}}{3^{\alpha+2}} p_1^2
+\frac{2^\alpha(1-\mu) -(2^\alpha-4) p_0}{3^{\alpha+1}} p_1 +\\ \notag
&+&
2p_0\frac{ -1 + \mu + p_0}{3^\alpha}, \\
p'_1&=&   
2\frac{2\, 3^\alpha + 2^\alpha -3 }{3^{\alpha+2}} p_1^2+
\\ \notag
&+&   
\frac{  (8 \,3^\alpha+2^\alpha )(\mu-1+p_0)  +
4-6\mu  
 -12 p_0 }{3^{\alpha+1}} p_1+
\\
&+& 2(1-\mu + [ 2/3^\alpha-1] p_0)( 1 -  \mu - p_0).
\end{eqnarray}

It is easy to see that, for any $\alpha$,
$(p_0,p_1)=(1-\mu,0)$ is a fixed point, 
and hence also $(p_0,p_1,p_2,p_3)=(1-\mu,0,0,\mu)$, as obtained after using the normalization and mean value conditions, 
$p_3=1-\sum_{j\neq3} p_j$ and $p_2=3-2p_1-3p_0-3\mu$. 
However, this solution is unstable (it contains as particular case the unstable solution found for $\mu=1/2$). 
To find the second, nontrivial and stable, fixed point, we turn to a numerical solution method. 
Let us note that, unlike the case $M=2$, for $M=3$ (or higher), asymmetric states with $\mu=1/2$ can exist. Notably, an initial condition with $\mu=1/2$ (even if asymmetric) will evolve towards the corresponding symmetric state, depending on the value of $\alpha$, as it is the only attractor with that mean. 

\begin{figure}[h!]
\includegraphics[width=0.7\columnwidth]{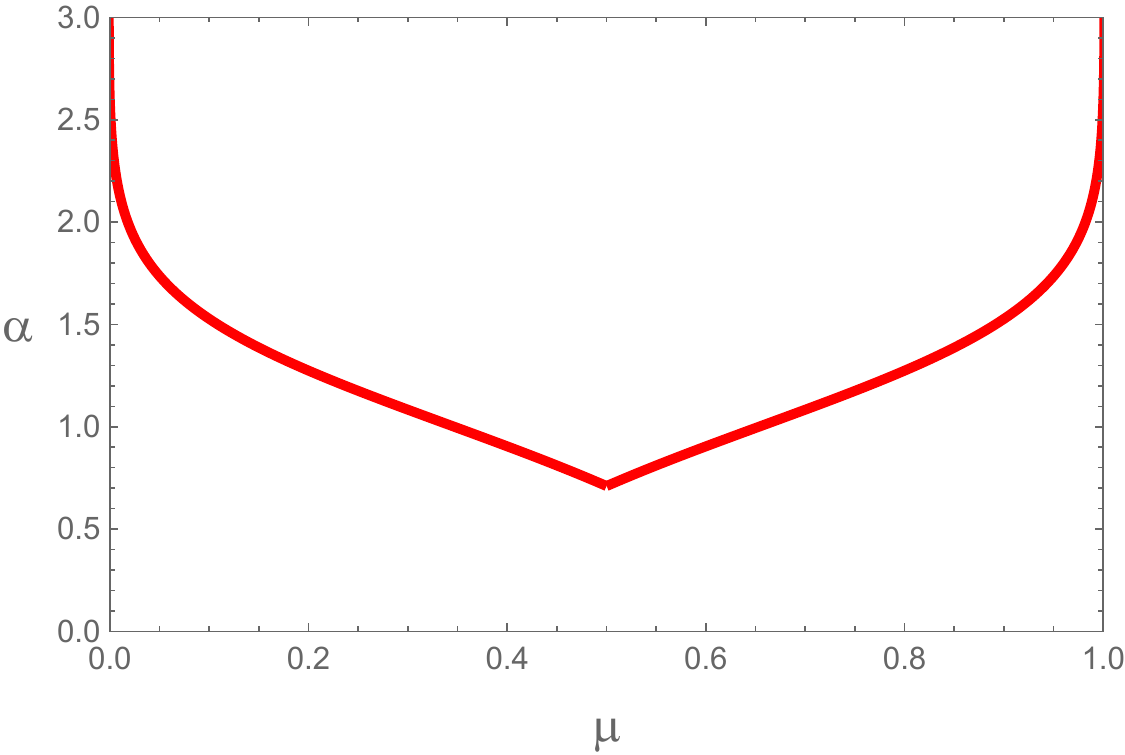}
    \caption{Critical curve $\alpha$ vs. $\mu$, for $M=3$, described by Eq.~(\ref{eq:critical}). It defines the boundary between bimodality (bipolarization) above the line and unimodal behavior (consensus) below it. The greater the asymmetry, the stronger the degree of homophily required to yield bipolarization.
 }
    \label{fig:kcritxmu}
\end{figure}

We observe that, for a given $\mu$, bimodality occurs when the homophily exponent $\alpha$ exceeds a critical value $\alpha_{c}$.
To estimate the critical line $\alpha$ vs. $\mu$ that separates the unimodal and bimodal regimes, we set the condition $p_0=p_1$ for $\mu>1/2$, in which case we obtain the inverse relation
\begin{eqnarray} \label{eq:critical}
\mu  &=& \frac{4\, 27^\alpha + (-6 + 2^\alpha)^2 - 3^\alpha (-6 + 2^\alpha) (4 + 2^\alpha)}{ 4\,27^\alpha + 6 (-6 + 2^\alpha)^2 - 3^\alpha (-36 + 4^\alpha)}.\;\;\;\;
\end{eqnarray}
Additionally, noting that the critical $\alpha$ must be symmetric around $\mu=1/2$, we obtain the full curve, which is plotted in Fig.~\ref{fig:kcritxmu}, with bipolarity occurring in the parameter region above the curve. 
Note that, the greater the asymmetry, the stronger the degree of homophily required to yield bipolarization.

\begin{figure}[b!]
\includegraphics[width=0.7\columnwidth]{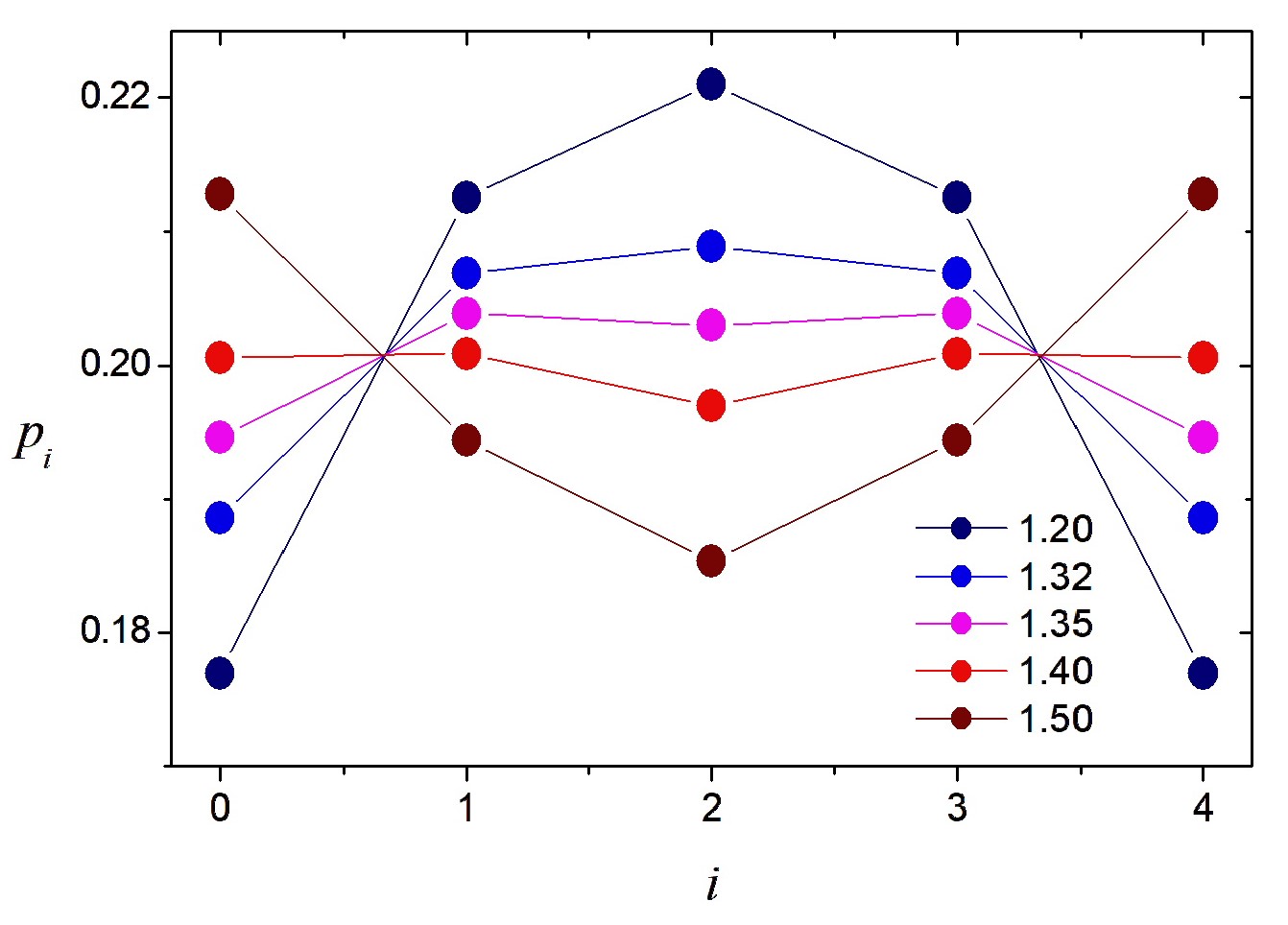}
\caption{Distribution $p_i$ vs $i$, for $M=4$ (symmetric case), and different values of $\alpha$ indicated in the legend.  
}
\label{fig:M4}
\end{figure}

\begin{figure*}[ht!]
\includegraphics[width=0.75\textwidth]{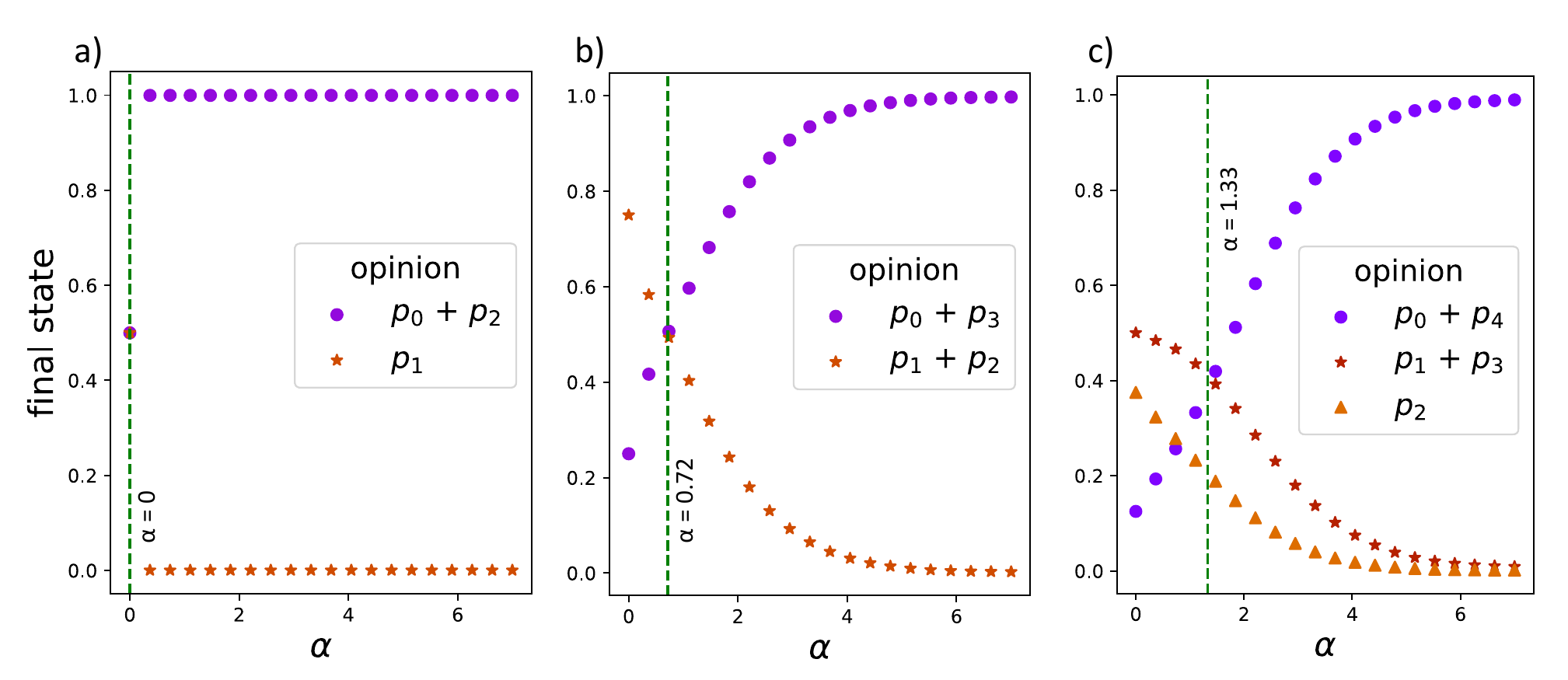}
\caption{For small values of 
$M$ (symmetric case), the final proportion of agents with each possible number of positive arguments is shown. The numerical solution of the rate equation (symbols) is compared with the analytical result for the transition, calculated in this section (dashed line), for $M=2$ (a) $M=3$ (b) and $M=4$ (c).
 }
\label{fig:allsmall}
\end{figure*}

\subsubsection{$M=4$}

For $M=4$ arguments, we will restrict the analysis to the symmetric case, for which holds $p_4=p_0$, $p_3=p_1$ and the normalization condition $p_2=1-2p_0-2p_1$, such the system is ruled by the following set of two equations 

\begin{eqnarray} \notag
 p'_0 &=&  4^\alpha (2^\alpha - 4\, 3^\alpha + 3\, 4^\alpha) p_1^2 + 2^{3(1+\alpha)} (2 p_0 -1) p_0 + \\ 
&& + 2^{2 \alpha+1} [3^\alpha +(- 6 + 2^{3+\alpha} - 2 \,3^\alpha) p_0 ] p_1,  \\ \notag
 p'_1 &=&   5\,2^\alpha (2^\alpha-1) p_1^2 + 
 2^{\alpha+3} [-2^\alpha + (2^{\alpha+1}-2) p_0] p_1 + \\
 && + 2^{\alpha+1}  (2 p_0 -1 ) [-2^\alpha + (2^{\alpha+1}-4)p_0]. 
\end{eqnarray}
Again in this case, it is evident that for any $\alpha$,
$(p_0,p_1)=(1/2,0)$ and therefore $(p_0,p_1,p_2,p_3,p_4)=(1/2,0,0,0,1/2)$ is a fixed point, although unstable.
The stable fixed point is obtained numerically.
Profiles of $p_i$ vs. $i$ are presented in Fig.~\ref{fig:M4}, for different values of the homophily exponent $\alpha$.
As can be seen in this figure, the critical value of $\alpha$ between unimodal and bimodal behaviors is 
$\alpha_c \simeq 1.33$.

Figure~\ref{fig:allsmall} summarizes the results obtained for small values of $M$, when the initial configuration is symmetric.

\subsection{With homophily: Large values of $M$}
\label{sec:large}

\subsubsection{Derivation of the limit equation.}

To analyze the system of rate equations for large values of $M$, we approximate it by a continuous equation.  
Consider the probability measure $f_t^M = \frac{1}{M+1}\sum_i p_i(t)\delta_{i/M}$. 
Then for any continuous observable $\phi:[0,1]\to \mathbb{R}$, we have 
$\int\phi \,df_t^M = \frac{1}{M+1} \sum_i p_i(t) \phi(i/M)$. 
Taking the derivative in time, using Eq.~\eqref{eq:RateEq}, and then performing a second-order Taylor expansion neglecting 
$O(1/M^2)$ terms leads to (see Appendix \ref{App_Cont_Eq}): 
\begin{equation}\label{ContEq20}
\begin{split}
\frac{d}{dt} \int_0^1\phi \,df_t^M 
=  \int_0^1 \phi'(x) v[f_t^M](x) df_t^M(x) \\
+  \int \phi''(x) D[f_t^M](x) df_t^M(x),
\end{split}
\end{equation}
for any $\phi$, with 
\begin{equation}\label{Def_vv}
 v[f](x) = (1+\frac1M)\int_0^1 h(x,y) (y-x)\,df(y) 
\end{equation}
and
\begin{equation}\label{Def_DD}
D[f](x) = \frac{1}{2M}\int_0^1 h(x,y) (y+x-2xy)\,df(y).  
\end{equation}
This is the weak formulation of 
\begin{equation}\label{FP}
  \partial_t f_t + \partial_x(v[f_t](x)f_t) =  \partial_{xx}(D[f_t](x)f_t) 
\end{equation}
with no-flux boundary conditions, where $v[f_t]$ is the drift and $D[f_t](x)$ the diffusion coefficient.

The transport term $\partial_x(v[f_t](x)f_t)$ tends to concentrate the distribution as a consequence of the compromise rule implemented in Eq.~\eqref{eq:update_rule}. On the other hand, the diffusion term tends to push the mass toward the boundary of the interval $[0,1]$. 
Bipolarization/non-polarization results from the balance between the intensity of diffusion, controlled by $M,$ and of the tendency to compromise, controlled by $\alpha$. 

Note that, as for the system of rate equations, Eq.~\eqref{ContEq20} 
conserves mass and the mean value. This can easily verified by taking $\phi(x)=1$ and $\phi(x)=x$ as test functions. 
Additionally, the equation is invariant under the transformation $x\to 1-x$. Specifically, if $f_t$ is a solution, then the measure $g_t$, 
defined by $\int \phi(x)\,dg_t(x):=\int \phi(1-x)\,df_t(x)$, also satisfies 
the same equation. This follows from the fact that $v[g_t](1-x)=-v[f_t](x)$ and $D[g_t](1-x)=D[f_t](x)$. Consequently, if $f_0$ is symmetric, then $f_t$ remains symmetric for all $t$.

\subsubsection{ $M=+\infty$}

In the limit $M\to +\infty$, the diffusion term in Eq.~\eqref{ContEq20} disappears and we get 
\begin{equation}
     \partial_t f_t + \partial_x(v[f_t](x)f_t) = 0, 
\end{equation}
where only the tendency to compromise can be felt, through the drift 
$ v[f](x) = \int_0^1 h(x,y) (y-x)\,df(y)$. 
As a consequence, $f_t$ converges to a Dirac delta as $t\to +\infty$ , which is necessarily located at $x=\mu$ since $\mu$ is conserved. 
This result is well-known when $\alpha=0$, but virtually the same proof holds for any $\alpha>0$, as only the symmetry of $h$ matters. 
Indeed taking $\phi(x)=x^2$ as test function gives 
\begin{equation}
   \frac{d}{dt} \frac12 \langle x^2\rangle = \iint h(x,y)x(y-x)\,df_t(x)df_t(y).  
\end{equation} 
Exchanging $x$ and $y$ and recalling that $h(x,y)=h(y,x)$, 
we also obtain
\begin{equation}
    \frac{d}{dt} \frac12 \langle x^2\rangle = -\iint h(x,y)y(y-x)\,df_t(x)df_t(y). 
\end{equation} 
Summing these two equations we obtain 
\begin{equation}\label{Eq100}
\frac{d}{dt}\langle x^2\rangle = -\iint h(x,y)(y-x)^2\,df_t(x)df_t(y). 
\end{equation}
Since $\mu$ is constant, the r.h.s. is the derivative of the variance of $f_t$, $Var[f_t]:=\langle x^2\rangle - \mu^2$, 
which thus decreases to $0$ - in particular, $f_t$ converges to a Dirac mass.

\subsubsection{ $M<\infty$}

Following the explicit computation performed above for constant homophily, which yields an approximate stationary state for the rate equation composed of a Gaussian and two Dirac masses at 0 and 1, 
we look for a stationary solution of the form 
\begin{equation}
    f_\infty = a\delta_0 + b\delta_1 + g(x)dx, 
\end{equation}
where $\delta_0$ and $\delta_1$ are Dirac masses at 0 and 1 with weights $a,b\ge 0$ (the final proportions of agents holding only positive or only negative arguments), and $g$ is a smooth function. 
Inserting this ansatz into Eq.~\eqref{ContEq20}, with the l.h.s. equal to 0, and integrating by parts , we eventually obtain the following equations 
for $a$, $b$ and $g$: 
\begin{equation}\label{FinalSystem10}
\begin{split}
& g(x) = g(0)\exp\Big(\int_0^x \frac{v[f_\infty](t)-D[f_\infty]'(t)}{D[f_\infty](t)}\,dt\Big) ,  \\
& a = \frac{g(0)}{2(M+1)},\qquad b = \frac{g(1)}{2(M+1)}, \\
& a+b+ \int_0^1 g(x)dx = 1, \\
& b + \int_0^1 xg(x)dx = \mu.
\end{split}
\end{equation}
Notice that the first two equations remain unchanged if $a,b,g$ are multiplied by a constant. 
The last two equations are normalization conditions corresponding to the conservation of the mass and of the mean value.

This system of equations can be simplified by assuming that $f_\infty$ is symmetric, i.e.  
$a=b=\frac{g(0)}{2(M+1)}$, and that
$g$ is symmetric in the sense that $g(x)=g(1-x)$.  
In particular, this implies that  $\mu=1/2$ and 
\begin{equation}
    \int_0^1 xg(x)dx = \int_0^1 (1-x)g(1-x)dx = \int_0^1 (1-x)g(x)dx ,
\end{equation}
so that $\int_0^1 xg(x)dx = \frac12 \int_0^1 g(x)dx$. 
The last two equations in Eq.~\eqref{FinalSystem10} are identical. 
Therefore, the symmetric stationary state is $f_\infty = a\delta_0 + b\delta_1 + g(x)dx$ with 
\begin{eqnarray}\label{SymmetricStatState1}
&& g(x) = g(0)\exp\Big(\int_0^x \frac{v[f_\infty](t)-D[f_\infty]'(t)}{D[f_\infty](t)}\,dt\Big), \\ \label{SymmetricStatState2}
&&\frac{g(0)}{M+1} + \int_0^1 g(x)dx = 1, \\
&& a = b = \frac{g(0)}{2(M+1)}. \label{SymmetricStatState3}
\end{eqnarray}

An explicit expression for $g$ can be derived in the case of constant homophily $h\equiv 1$ with $\mu=1/2$
Indeed, in that case, it follows from Eqs.~\eqref{Def_vv} and \eqref{Def_DD} that 
$$ v[f](t) = (1+\frac1M) (\frac12 - t), \quad D[f](t) = \frac{1}{4M}, \quad D[f]'(t) = 0. $$ 
Inserting these expressions into Eq.~\eqref{SymmetricStatState1} gives 
$$ g(x) =  g(0) \exp\Big( 4(1+M) \int_0^x (1/2- t)\,dt \Big) = C e^{-2M(1/2 - x)^2 }, $$ 
where the constant $C$ is such that the normalization condition, Eq.~ \eqref{SymmetricStatState2} holds. 
For $M\gg 1$, 
$$ \int_0^1 g(x)dx \approx \int_{-\infty}^\infty g(x)dx = C \int_{-\infty}^\infty e^{-2Mx^2}dx
= C\sqrt{\frac{\pi}{2M}}.
$$ 
The normalization condition \eqref{SymmetricStatState2} thus gives 
$$ C = \Big(\frac{e^{-M/2}}{M+1} + \sqrt{\frac{\pi}{2M}}\Big)^{-1} \approx \sqrt{\frac{2M}{\pi}}.  $$
Therefore, the symmetric stationary state $f_\infty$ is well-approximated by 
 $$ f_\infty \approx \sqrt{\frac{2M}{\pi}} e^{-2M(1/2 - x)^2}dx + \frac{e^{-M/2}}{\sqrt{2\pi M}}(\delta_0+\delta_1) ,
 $$ 
which coincides with the symmetric stationary solution, in Eqs.~\ref{DefGaussianApprox}-\ref{DefDiracApprox2} found above.

For non-constant homophily, we solve numerically Eqs.~\eqref{SymmetricStatState1}-\eqref{SymmetricStatState3}
first introducing the scaled variables $(\tilde g, \tilde a):=(g/g(0), a/g(0))$, so that $\tilde a = \frac{1}{2(M+1)}$ and 
\begin{equation}\label{Equtilde}
\tilde g(x) = \exp\Big(\int_0^x \frac{v(t)-D'(t)}{D(t)}\,dt\Big) ,
\end{equation}
where 
\begin{eqnarray*}
 \frac{M}{M+1}v(x) & = & -\tilde axh(x,0) + \tilde a h(x,1)(1-x) \\
&& + \int_0^1 h(x,y) (y-x)\tilde g(y)\,dy, 
\end{eqnarray*}
and
\begin{eqnarray*}
 2M D(x) &= &\tilde a xh(x,0) + \tilde a h(x,1)(1-x) \\
&& + \int_0^1 h(x,y) (y+x-2xy)\tilde g(y)dy.  
\end{eqnarray*}
Equation \eqref{Equtilde} was solved using scipy.fsolve while enforcing the symmetry of $\tilde g$. 
Then $\tilde g$ was normalized to ensure Eq.~\eqref{SymmetricStatState2}, obtaining 
$g=\tilde g/(\frac{1}{M+1}+\int_0^1 \tilde g(x)dx)$. 
The Dirac weights $a$ and $b$ are then given by Eq.~\eqref{SymmetricStatState3}.

\begin{figure}[h!]
\includegraphics[width=0.48\textwidth]{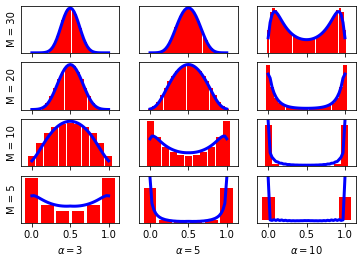}
\caption{Stationary state of the rate equations (red) and continuous solution (blue) for $M=5,10,20,30$ (bottom to top)
and $\alpha=3,5,10$ (left to right). 
}
\label{Fig:cont_vs_rate}
\end{figure}

Figure \ref{Fig:cont_vs_rate} shows the continuous approximation $g$ (blue) without the Dirac masses $a(\delta_0+\delta_1)$ 
together with the solution of the rate equations (red) for $M=5,10,20,30$ and $\alpha=3,5,10$, the same as in Fig.~\ref{fig:Rate_eq}. 
As expected, the agreement between both solutions improves as $M$ increases, 
although the presence of Dirac masses at 0 and 1 for large $\alpha$ slows down the convergence.   
Nonetheless, the continuous approximation still captures the shape of the stationary state very well.

\section{Conclusions}

In this work, we proposed a minimal model that produces bipolarization without imposing negative influences or nonlinear interactions. By deriving rate equations, whose solutions are in full agreement with agent-based simulations in complete graphs, we managed to systematically investigate the impact of three key parameters: the number of arguments $M$, the mean value $\mu$ of the distribution of the proportion of positive arguments, and the exponent $\alpha$ that controls the strength of homophily. 

Exact results were derived for the case without homophily and, in general, for small values of $M$ with homophily. In the former case, the system always reaches consensus, highlighting the necessity of homophily for achieving bipolarization. 

With homophily and for large values of $M$, we derived a continuous equation that shows the transition between consensus and bipolarization arises from the competition between homophily and interaction noise. Stronger homophily promotes bipolarization, while a higher number of arguments favors consensus. In the limit $M \to \infty$, 
 only consensus is possible.

 Therefore, while simple, this minimal model based on PAT framework offers valuable insights into the roles of homophily and the number of arguments.

 As a natural extension, implementing this dynamics in complex networks would be an interesting avenue for further exploration.

{\bf Acknowledgments:}
C.A. acknowledges partial financial support received from
 Conselho Nacional de Desenvolvimento Científico e Tecnológico (CNPq)-Brazil (311435/2020-3), Fundação de Amparo à Pesquisa do
Estado de Rio de Janeiro (FAPERJ)-Brazil (CNE E-26/204.130/2024), and Coordenação de Aperfeiçoamento de Pessoal de Nível Superior (CAPES)-Brazil (code 001).
P.B and L.P acknowledge the support of the Universidad de Buenos Aires (UBA) through Grant UBACyT, 20020220100181BA and the Agencia Nacional de Promoción de la Investigación, el Desarrollo Tecnológico y la Innovación through Grant No. PICT-2020-SERIEA-00966. 
J.P.P. and N.S. acknowledge the support of the Consejo Nacional de Investigaciones Científicas y Tecnológicas (CONICET) and Universidad de Buenos Aires (UBA) by grants PIP 11220200102851CO and UBACYT 20020170100445BA.

 \appendix

 \section{Derivation of the continuous Eq.~\eqref{ContEq20}}
 \label{App_Cont_Eq}

The system of rate equations can be rewritten as a single equation for the probability measure 
$f_t^M = \frac{1}{M+1}\sum_i p_i(t)\delta_{i/M}$ as follows. 
For any continuously differentiable function $\phi:[0,1]\to \mathbb{R}$, we have 
$\int\phi \,df_t^M = \frac{1}{M+1} \sum_i p_i(t) \phi(i/M)$. 
Taking the derivative in time, using Eq.~\eqref{eq:RateEq}, gives 
$$ \frac{d}{dt} \int\phi \,df_t^M = \frac{1}{M+1} \sum_i p_i'(t) \phi(i/M) = A+B-C,  $$
with 
\begin{align*}
 A & = \frac{1}{M+1} \sum_{i=0}^M p_{i+1}\sum_j p_j h(i+1,j)\frac{i+1}{M}\frac{M-j}{M} \phi(i/M) \\
 & = \frac{1}{M+1} \sum_{i,j=0}^M p_i p_j h(i,j)\frac{i}{M}\frac{M-j}{M} \phi((i-1)/M),  
\end{align*}
\begin{align*}
B & = \frac{1}{M+1} \sum_{i,j=0}^M p_{i-1} p_j h(i-1,j)\frac{M-(i-1)}{M}\frac{j}{M} \phi(i/M)\\
 & = \frac{1}{M+1} \sum_{i,j=0}^M p_i p_j h(i,j)\frac{M-i}{M}\frac{j}{M} \phi((i+1)/M),
\end{align*}
$$ C  = \frac{1}{M+1} \sum_{i,j=0}^M p_i p_j h(i,j) 
\Big( \frac{i}{M}\frac{M-j}{M} + \frac{M-i}{M}\frac{j}{M} \Big)\phi(i/M). $$
Thus 
$$ \frac{d}{dt} \int\phi \,df_t^M = C-D, $$
with 
$$ C = \frac{1}{M+1} \sum_{i,j=0}^M p_i p_j h(i,j)\frac{M-i}{M}\frac{j}{M} \Big(\phi((i+1)/M)-\phi(i/M)\Big), $$
and 
$$ D = \frac{1}{M+1} \sum_{i,j=0}^M p_i p_j h(i,j)\frac{i}{M}\frac{M-j}{M} \Big(\phi(i/M)-\phi((i-1)/M)\Big).$$ 
Taylor expanding up to the second order yields
\begin{align*}
& M(M+1)\frac{d}{dt} \int\phi \,df_t^M = \\
&   \sum_{i,j=0}^M p_i p_j h(i,j) 
    \{(\alpha-\beta) \phi'(i/M)
          + \frac{1}{2M} (\alpha+\beta)\phi''(i/M)\} \, ,
\end{align*}
with $\alpha = \frac{j}{M}\frac{M-i}{M}$, $\beta=\frac{i}{M}\frac{M-j}{M}$. 
The sums can be rewritten as integral with respect to $f_t^M$: 
\begin{align*}
& \frac{M}{M+1} \frac{d}{dt} \int\phi \,df_t^M = \\
 &  \iint h(x,y)\phi'(x) (y(1-x)-x(1-y))\,df_t^M(x) df_t^M(y) \\
& + \frac{1}{2M} \iint h(x,y)\phi''(x) (y(1-x)+x(1-y))\,df_t^M(x) df_t^M(y),
\end{align*}
plus a term $O(1/M^2)$ that we neglect. Then we deduce Eq.~\eqref{ContEq20}.

\section{Stationary state of the continuous equation}
\label{App_Stat_state}

We look for a stationary solution of Eq.~\eqref{ContEq20} of the form 
$$ f_\infty = a\delta_0 + b\delta_1 + g(x)dx, $$ 
for some non-negative weights $a,b\ge 0$ and some smooth nonnegative function $g$.  
Inserting this ansatz in Eq.~\eqref{ContEq20} with the l.h.s. equal to 0 gives 
$$ \int_0^1 \phi'(x) v(x) df_\infty(x) + \int \phi''(x) D(x) df_\infty(x) = 0, $$
where we let $v=v[f_\infty]$ and $D:=D[f_\infty]$ for ease of notation. 
This can be rewritten explicitly in terms of $a,b,g$ as 
\begin{eqnarray*}
&& a\phi'(0) v(0) + b\phi'(1) v(1)  +  a \phi''(0) D(0) +  b \phi''(1) D(1) \\
&& + \int_0^1 \phi'(x) v(x) g(x)dx
+  \int_0^1 \phi''(x) D(x) g(x)dx = 0. 
\end{eqnarray*}
Integrating by parts in the integrals, we obtain :
\begin{equation}\label{App_Stat_State_100}
\begin{split}
 0 = & \phi'(0)\Big(av(0) - (Dg)(0)\Big) + \phi'(1)\Big(bv(1) + (Dg)(1)\Big) \\
   & + \phi(1)\Big(v(1)g(1) - (Dg)'(1)\Big) \\
   & + \phi(0)\Big(-v(0)g(0) + (Dg)'(0)\Big) \\
   & + \int_0^1 \phi(x)\Big( (Dg)''(x) - (vg)'(x)\Big)\,dx,
\end{split}
\end{equation}
which holds for any $\phi$. 

Note that we neglected the terms involving $\phi''(0)$ and $\phi''(0)$.  
Indeed, Eq.~\eqref{ContEq20} was derived by neglecting terms in $O(1/M^2)$, which would involve $\phi^{(3)}$. 
When integrating by parts, these terms would yield contributions involving $\phi''(0)$ and $\phi''(1)$, which are not taken into account in the current approximation.  

We thus deduce from Eq.~\eqref{App_Stat_State_100} that 
\begin{eqnarray}
&& (Dg)''(x) - (vg)'(x) = 0. \label{App_Stat_state_1} \\ 
&& v(0)g(0) = (Dg)'(0),\, v(1)g(1) = (Dg)'(1) \label{App_Stat_state_2}  
\end{eqnarray}
and
\begin{equation}\label{App_Stat_state_3} 
av(0) = (Dg)(0),\, bv(1) = - (Dg)(1).  
\end{equation}
Integrating Eq.~\eqref{App_Stat_state_1}, Eq.~ using \eqref{App_Stat_state_2}, gives $(Dg)'(x) - (vg)(x) = 0$, 
i.e., $g'(x) = \frac{v-D'}{D} g$, 
or 
\begin{equation}\label{App_Stat_state_6} 
 g(x) = g(0)\exp\Big(\int_0^x \frac{v[f_\infty](t)-D[f_\infty]'(t)}{D[f_\infty](t)}\,dt\Big).  
\end{equation}
Moreover, by the definition of $v$ and $D$, in Eqs.~\eqref{Def_vv} and \eqref{Def_DD} respectively,  
$$ \frac{M}{M+1}v(0) = 2MD(0) = \int_0^1 h(0,y) y\,df_\infty(y), $$
and
$$ \frac{M}{M+1}v(1) = -2MD(1) = \int_0^1 h(1,y) (y-1)\,df_\infty(y). $$
Thus Eq.~\eqref{App_Stat_state_3} gives 
\begin{equation}\label{App_Stat_state_7} 
 a = \frac{g(0)}{2(M+1)},\qquad 
b = \frac{g(1)}{2(M+1)}. 
\end{equation}
Equations \eqref{App_Stat_state_6} and \eqref{App_Stat_state_7} are the first two equations in the system defined in Eq.~\eqref{FinalSystem10}.

As a final comment, notice that if we had insisted on taking into account the terms 
$\frac{am_0}{2M}\phi''(0) + \frac{b(1-m_0)}{2M} \phi''(1)$, 
we would have obtained $a=b=0$, which leads to $g(0)=g(1)=0$, and thus $g\equiv 0$. 
This implies that $f_\infty=0$, which is absurd, since $f_\infty$ has unit total mass. 
\\

\end{document}